\documentclass[a4paper,twocolumn,twoside,10pt]{article}

\usepackage{amssymb}
\usepackage{graphicx}
\usepackage{abstract}
\usepackage{url}

\usepackage{mathptmx} 

\usepackage{geometry}
\usepackage{fancyhdr}
\usepackage{titlesec}
\titlelabel{\thetitle.\quad}
\titleformat*{\section}{\bfseries}
\titleformat*{\subsection}{\slshape}
\titleformat*{\subsubsection}{\itshape}

\usepackage[font=small,format=plain,labelfont=bf]{caption}

\makeatletter
\renewcommand\@makefntext[1]{\leftskip=0.6em\hskip-0.5em\@makefnmark#1}
\makeatother

\def\Mo{{M_{0}}}

\def\Md{{M_{\rm dyn}}}
\def\Msol{{M_{\odot}}}
\def\Mstar{{M_{\star}}}
\def\Lsol{{L_{\odot}}}
\def\am{{a_{\rm M}}}

\def\ac{{a_{\rm c}}}
\def\gn{{g_{\rm N}}}
\def\vc{{\upsilon_{\rm c}}}

\def\mss{{\,m\,s$^{-2}$}}
\def\L{{{\mathcal L}}}
\def\R{{{\mathcal R}}}
\newcommand\vol[1]{{\bf #1}}
\newcommand\name[1]{{\small\sc #1}}
\newcommand\para[1]{{\vspace{1mm}\noindent{\bf #1}}}
\newcommand\epara[1]{{\vspace{1mm}\noindent{\it #1}}}

\author{ 
\normalsize Sascha Trippe \\
\small Department of Physics and Astronomy, Seoul National University, Gwanak-gu, Seoul 151-742, South Korea \vspace{-1mm} \\
\small E-mail: trippe@astro.snu.ac.kr
}

\title{ 
\linespread{0.85}\vspace{1mm}\LARGE\bf{The ``Missing Mass Problem'' in Astronomy and the Need for a Modified Law of Gravity}
}

\date{\small Z. Naturforsch. A (in press) --- received 25 July 2013; revised 15 January 2014; accepted 17 January 2014}

\begin{document}

\twocolumn[  
\maketitle
\begin{onecolabstract}
\noindent\emph{Abstract:} Since the 1930s, astronomical observations have accumulated evidence that our understanding of the dynamics of galaxies and groups of galaxies is grossly incomplete: assuming the validity of Newton's law of gravity on astronomical scales, the observed mass (stored in stars and interstellar gas) of stellar systems can account only for roughly 10\% of the dynamical (gravitating) mass required to explain the high velocities of stars in those systems. The standard approach to this ``missing mass problem'' has been the postulate of ``dark matter'', meaning an additional, electromagnetically dark, matter component that provides the missing mass. However, direct observational evidence for dark matter has not been found to date. More importantly, astronomical observations obtained during the last decade indicate that dark matter cannot explain the kinematics of galaxies. Multiple observations show that the discrepancy between observed and dynamical mass is a function of gravitational acceleration (or field strength) but not of other parameters (size, rotation speed, etc.) of a galaxy; the mass discrepancy appears below a characteristic and universal acceleration $\am=(1.1\pm0.1)\times10^{-10}$\mss\ (``Milgrom's constant''). Consequently, the idea of a modified law of gravity, specifically the ansatz of \emph{Modified Newtonian Dynamics} (MOND), is becoming increasingly important in astrophysics. MOND has successfully predicted various important empirical relations of galaxy dynamics, including the famous Tully--Fisher and Faber--Jackson relations. MOND is found to be consistent with stellar dynamics from binary stars to clusters of galaxies, thus covering stellar systems spanning eight orders of magnitude in size and 14 orders of magnitude in mass. These developments have the potential to initiate a paradigm shift from dark matter to a modified law of gravity as the physical mechanism behind the missing mass problem.
\vspace{1.5mm} 
\end{onecolabstract}
]

\section{Introduction \label{sect_intro}}

A \emph{missing mass problem} in astronomy was noted for the first time by Fritz Zwicky as early as 1933 \cite{zwicky1933}. From observations of the Coma cluster of galaxies he estimated the potential energy stored in the cluster like

\begin{equation}
\label{eq_comaenergy}
W = -\frac{3\,G\,M^2}{5\,R} = -M\,\sigma^2
\end{equation}

\noindent
with $G$ being Newton's constant, $R$ denoting the radius of the cluster, $M$ being the total mass of the cluster, and $\sigma$ denoting the (three-dimensional) velocity dispersion, i.e., the root-mean-square speed (about the mean) of the cluster galaxies. The first equality arises from approximating the galaxy cluster as a homogeneous sphere with constant mass density; the second equality is given by the virial theorem (e.g., \cite{binney2008}). Images of the cluster provided estimates for the radius and the observable \emph{luminous} mass $\Mo$ stored in the stars of the individual galaxies; spectroscopic observations provided the galactic line-of-sight velocities and thus $\sigma$. Surprisingly, \cite{zwicky1933} had to conclude that the observed value for $\sigma$ ($\approx1500$\,km\,s$^{-1}$) exceeded the one expected by a factor of about 20; because $M\propto\sigma^2$, this indicated a \emph{dynamical}, gravitating, mass of the system $\Md\approx400\Mo$. This obvious discrepancy led Zwicky to the conjecture that most of the mass of galaxy clusters is non-luminous and thus not observable at optical wavelengths, eventually making him coin the term \emph{dark matter}. Modern observations find that the mass discrepancy is far less severe than feared by Zwicky: the overwhelming fraction of the luminous mass of galaxy clusters is indeed not stored in stars but in the diffuse intra-cluster medium -- hot gas with virial temperatures on the order of millions of Kelvin and observable only at X-ray wavelengths. However, even when adding up all luminous matter, a substantial mass discrepancy remains: for an average galaxy cluster, $\Md/\Mo\approx8$ \cite{giodini2009}.

Since the 1970s it has become clear that not only groups of galaxies but also individual galaxies show a discrepancy between luminous and dynamical mass. The most evident kinematic signature is provided by the rotation curves -- meaning the circular speed $\vc$ as function of radius $r$ from the galaxy center -- of disk galaxies. In standard Newtonian dynamics and assuming circular orbits \cite{binney2008},

\begin{equation}
\label{eq_galrot}
\vc^2(r) = r\,\frac{\partial\Phi}{\partial r} \approx \frac{G\,M(r)}{r}
\end{equation}

\noindent
with $\Phi$ denoting the gravitational potential and $M(r)$ being the mass enclosed within $r$; the second equality becomes exact in the case of a homogeneous spherical mass distribution. In the inner regions of galaxies, $M(r)$ can be derived from integration over the mass density as function of radius $\rho(r)$, which can result in a complicated profile depending on the actual galaxy. However, in the outer regions of galaxies where the mass density is low, we expect $M(r)\approx\Mo\approx const$ and, accordingly, $\vc\propto1/\sqrt{r}$ -- the circular speed should decrease with increasing radial distance from the center of the galaxy. The observations find a completely different result: the outer regions of disk galaxies obey the law $\vc\approx const$ -- the rotation curves become flat \cite{rubin1980}. The interpretation that $\Md>\Mo$, by up to one order of magnitude, is further supported by arguments with respect to the stability of galactic disks \cite{ostriker1973,gallagher1976}. 

Yet another indication toward a lack of observed mass is provided by cosmology. Modern cosmological observations imply a certain fixed value for the density of matter/energy within the universe. A combination of theoretical predictions and the dynamical observations discussed above\footnote{And ignoring here an additional component, \emph{dark energy}, an electromagnetically dark fluid generating a negative pressure that has been postulated as explanation for the accelerated expansion of the universe.} leads to the conclusion that only about 16\% of the matter present in the universe is actually luminous -- on cosmological scales, we observe a mass discrepancy $\Md/\Mo\approx6$ \cite{bahcall1999,ade2013}.

Evidently, the missing mass problem is a substantial challenge not only for astronomy but for various fields of physics. In the following, I introduce the two main concepts proposed for solving the missing mass problem, \emph{non-baryonic dark matter} on the one hand and \emph{modified laws of gravity} on the other hand. I present and discuss the increasing body of observational evidence in favor of modified gravity that has been accumulated especially within the last decade. Eventually, I argue that these recent observations have the potential to initiate a paradigm shift in astrophysics from theories based on dark matter toward theories based on modifications of the laws of gravity on astronomical scales.

\section{Dark Matter \label{sect_dm}}

The most evident approach to the missing mass problem is the assumption that our observations are incomplete and that the mass discrepancy arises from additional matter components simply not yet observed. A first clue toward the distribution of this additional dark mass is provided by the rotation of galaxies:  the rotation curves behave as if galaxies were surrounded by halos of matter extending well beyond the visible components of the galaxies. As one can estimate from (\ref{eq_galrot}) easily, $\vc\approx const$ indicates halos with density profiles $\rho(r)\propto r^{-2}$. Technically, a simple parameterization of a dark matter halo is achieved by a mass density profile $\rho(r) = \rho_0 (r_0/r)^2$, with $\rho_0$ being a scaling factor of the dimension of a mass density and $r_0$ denoting a characteristic radius.\footnote{Such a functional form of the density profile follows naturally from the assumption that a dark matter halo is a self-gravitating isothermal ensemble of particles in equilibrium; cf. \S\,4.3.3(b) of \cite{binney2008}. Modern studies use more sophisticated dark matter density profiles, especially the \emph{Navarro--Frenk--White profile} \cite{navarro1995}. All those profiles require at least as many parameters -- i.e., three -- as the simple powerlaw profile given here.} From this follows that any dark matter halo model for a given galaxy requires (at least) three free parameters: the scaling parameters $\rho_0$ and $r_0$ plus the galaxy's \emph{mass-to-light ratio} $\Upsilon$ which is needed to estimate the luminous mass of a galaxy from its brightness. This ratio is a function of the composition of the stellar population that makes up the galaxy.\footnote{The mass-to-light ratio is usually quoted in units of solar masses per solar luminosity, $\Msol/\Lsol$; for the inner regions of the Milky Way, $\Upsilon\approx3\,\Msol/\Lsol$ (e.g., \cite{binney2008}).}

Historically, the concept of dark matter evolved in three main steps.

\subsection{Interstellar matter \label{ssect_ism}}

A notable (crudely 10--50\% depending on the actual galaxy) fraction of the total -- luminous -- mass of galaxies is contributed not by stars but by the diffuse \emph{interstellar medium} (ISM), gas and dust distributed between the stars. Accordingly, it is tempting to identify the missing mass with additional ISM, and indeed ISM distributed throughout galaxy clusters largely explains the enormous mass discrepancies first reported by \cite{zwicky1933}. However, else than at the time of Zwicky, modern astronomical observations cover the entire electromagnetic spectrum from radio to $\gamma$-rays and are able to trace the characteristic signatures of interstellar matter. Atomic gas -- mostly hydrogen -- is identified via radio observations of the H\,{\sc i} 21-cm line. Hot, ionized gas is traced via its thermal X-ray emission. Molecular gas and dust show emission lines at radio and infrared wavelengths and are sources of a characteristic wavelength-dependent absorption of light emitted by background sources (\emph{interstellar extinction}). And, finally, interstellar dust is an emitter of blackbody radiation peaking in the mid to far infrared depending on the dust temperature (for an exhaustive review of interstellar matter, see, e.g., \cite{kwok2007}). Modern observations exclude the possibility that substantial amounts of diffuse interstellar matter have been missed within or around galaxies (cf. also \cite{gallagher1976}). The situation is somewhat more complex for the case of galaxy clusters: here the observational limits are less well established and could lead to corrections of $\Mo$ by factors on the order of two \cite{shull2012}.

\subsection{MACHOs \label{ssect_macho}}

In the 1980s, the absence of sufficient amounts of \emph{diffuse} interstellar matter led to the idea that the missing mass might be stored in small \emph{compact} bodies -- dubbed \emph{massive compact halo objects}, or MACHOs -- like interstellar planets or stellar mass black holes distributed within and around galaxies. Such objects are extremely difficult to detect as both the emission and the absorption of light by them would be marginal. However, the compactness of those bodies makes them suitable as gravitational lenses \cite{paczynski1986}: a MACHO located within or around the Milky Way focuses the light of a background star toward an observer on Earth, leading to a substantial (by factors $\approx2-10$) and characteristic increase of the observed brightness of the star over a time of a few hundred days. Accordingly, simultaneous long-term monitoring of large numbers of background stars should unveil the presence of MACHOs. However, several studies observing the Magellanic Clouds -- satellite galaxies of the Milky Way that provide millions of background stars -- as well as the Galactic spiral arms found only few gravitational lensing events. In conclusion, MACHOs contribute only few per cent of the missing mass of the Milky Way at most \cite{alcock2001,derue2001,wyrz2011a,wyrz2011b}.

\subsection{WIMPs \label{ssect_wimp}}

With interstellar matter -- diffuse or compact -- largely being ruled out, the search for dark matter candidates eventually reached the realm of particle physics by postulating the universal presence of non-baryonic \emph{weakly interacting massive particles}, or WIMPs for short. Massive numerical simulations (e.g., \cite{springel2005}) show that consistency with cosmic structure formation requires dark matter to be \emph{cold}, meaning that dark matter particles move at non-relativistic speeds (this excludes massive neutrinos as WIMP candidates -- they would be \emph{hot} dark matter moving at relativistic speeds).

Theoretically, the most ``popular'' source of cold dark matter particles is the \emph{supersymmetric sector} (SUSY) of particle physics \cite{wess1974,martin2011}. Supersymmetric particles are supposed to be more massive than standard model (SM) particles. Furthermore, a conserved quantum number -- the \emph{R-parity} -- ensures that SUSY particles may decay into lighter SUSY particles but not into SM particles. Accordingly, one may tentatively identify the WIMP with the lightest supersymmetric particle; usually, this is supposed to be either a \emph{gravitino} or a \emph{neutralino}. However, despite massive experimental efforts, observational evidence for a supersymmetric sector has not been found yet \cite{aad2013}.

\section{Modified Gravity \label{sect_mog}}

\begin{quote}
\small
{\bf ``}\,It is worth remembering that all of the discussion [on dark matter] so far has been based on the premise that Newtonian gravity and general relativity are correct on large scales. In fact, there is little or no direct evidence that conventional theories of gravity are correct on scales much larger than a parsec or so.\,{\bf ''}

--- \cite{binney1987}, \S\,10.4.3, p.\,635
\end{quote}

\noindent
Despite the overwhelming success of Einstein's General Theory of Relativity (GR) \cite{einstein1916}, modified laws of gravity can be, and have been, considered as solutions of the missing mass problem. As indicated by the quote from \cite{binney1987} given above, those proposals are motivated by the fact that GR is experimentally well-established in the regime of small spatial scales and strong gravitational fields (e.g., \cite{kramer2006}) but cannot be tested directly on spatial scales much larger than, and for gravitational fields much weaker than within, the solar system. Accordingly, modifications of Newton's law of gravity on spatial scales much larger than the scale of the solar system cannot be ruled out a priori.

The most obvious modification one can make is a change of the scaling of gravitational acceleration $g$ with radial distance $r$ for a point mass with luminous mass $\Mo$.\footnote{For simplicity, I only quote absolute values of positions, velocities, and accelerations.} In standard gravity, $g=\gn=G\Mo/r^2$, with $\gn$ denoting the standard Newtonian gravitational field strength. For a test particle orbiting $\Mo$ with circular speed $\vc$, the centripetal acceleration is $\ac=\vc^2/r\equiv g$, resulting immediately in $\vc^2=G\Mo/r$. We now assume a modified law of gravity $g=\eta(r/r_0)\,\gn$ with $r_0$ being a constant characteristic length, $\eta(r/r_0)=1$ for $r\ll r_0$, and $\eta(r/r_0)=r/r_0$ for $r\gg r_0$. The first limiting case corresponds to the usual Newtonian law of gravity; the second limiting case however results in $g=G\Mo/(rr_0)$ and, consequently, $\vc^2=G\Mo/r_0=const$ -- we have found a way to create constant circular speeds in the outer regions of disk galaxies. Our simple law of gravity is easily falsified however: we postulate here a modification at a characteristic length scale $r_0$, meaning that the mass discrepancy $\Md/\Mo$ for galaxies should correlate with the distance from the galactic center -- which is ruled out by observations \cite{mcgaugh2004}.\footnote{This is actually a non-trivial statement because for force laws deviating from a $r^{-2}$ scaling, Gauss' theorem is invalid (cf. \S\,2.1 of \cite{binney2008}). This implies that a test mass is influenced also by the mass distribution \emph{outside} its orbit. Fortunately, nature is on our side: the most important probes of stellar dynamics are disk galaxies. To good approximation, all disk galaxies can be parameterized by exponential surface mass density profiles. This means that for all disk galaxies the galaxy mass outside a given orbit is determined by the mass inside the orbit via the same functional relation (cf. also \S\,\ref{ssect_sda}). Similar, general parameterizations also exist for the distribution of mass in elliptical galaxies \cite{binney2008}.} Notable examples for those modified laws of gravity are given by \cite{finzi1963,sanders1984}.

A more sophisticated proposal was provided in 1983 by Mordehai Milgrom \cite{milgrom1983a,milgrom1983b,milgrom1983c}. At the time of Milgrom's work, the missing mass problem in galaxies was a fairly recent discovery, and the data base was still sparse. Observational features known to Milgrom were the asymptotically flat rotation curves of disk galaxies and the \emph{Tully--Fisher relation} \cite{tully1977} which suggested that $\Mo\propto\vc^4$ for disk galaxies (the Tully--Fisher relation will be discussed in detail in \S\,\ref{ssect_tully}). This led Milgrom to conjecture a modified law of gravity\footnote{Milgrom actually formulated his relation initially as a modified law of inertia but abandoned this interpretation quickly.} related to a characteristic \emph{acceleration} (or gravitational field strength) such that

\begin{equation}
\label{eq_mu}
\gn = \mu(x)\,g
\end{equation}

\noindent
with $x=g/\am$, $\am$ being a constant of the dimension of an acceleration -- today known as \emph{Milgrom's constant} -- and $\mu(x)$ being a transition function with the asymptotic behavior $\mu(x)=1$ for $x\gg1$ and $\mu(x)=x$ for $x\ll1$. Obviously, the first limiting case corresponds to the usual Newtonian dynamics. The second limiting case however results in

\begin{equation}
\label{eq_vc}
\vc^4 = G\,\Mo\,\am = const
\end{equation}

\noindent
for $g\ll\am$. By construction, (\ref{eq_vc}) assumes ordered circular motion of stars (or any test mass), corresponding to \emph{dynamically cold} stellar systems\footnote{This nomenclature stems from analogy to kinetic gas theory: a system is ``cold'' (``hot'') if the random velocities of the constituent particles are much lower (much higher) than any ordered, streaming velocity within the system.} like disk galaxies. For stellar systems dominated by random motions, i.e., \emph{dynamically hot} systems like elliptical galaxies and galaxy clusters, one finds \cite{milgrom1984,milgrom1994} -- again for $g\ll\am$ -- the relation

\begin{equation}
\label{eq_sigma}
\sigma^4 = \frac{4}{9}\,G\,\Mo\,\am = const
\end{equation}

\noindent
with $\sigma$ being the (three-dimensional) velocity dispersion. The proposed modified law of gravity (\ref{eq_mu}) and the resulting relations between luminous mass and velocity (\ref{eq_vc}, \ref{eq_sigma}) have been summarized under the term \emph{Modified Newtonian Dynamics}, or MOND for short.

The MOND laws provide a variety of explicit predictions (P) that put them in contrast to dynamics based on dark matter and that can be tested observationally:

\para{P1.} Like for \emph{any} modified law of gravity, the source of the gravitational field can only be the luminous source mass $\Mo$. This implies a one-to-one correspondence between the spatial structure of the gravitational field and the spatial distribution of mass. This is sharply distinct from the assumption of gravity being partially caused by dark matter: in this case, luminous and dark mass components may be spatially separate, resulting in a gravitational field not following the distribution of the luminous mass.

\para{P2.} According to (\ref{eq_mu}), MOND requires a \emph{universal} scaling of the mass discrepancy $\Md/\Mo$ (i.e., $g/\gn$) with acceleration, with the scaling law being the function $\mu(g/\am)$. This scaling law comes with only one free parameter, Milgrom's constant $\am$, which is a constant of nature.

\para{P3.} In the limit $g\ll\am$, stellar speeds necessarily scale like $\vc^4\propto\Mo$ and $\sigma^4\propto\Mo$ for ordered circular and random motions, respectively. Those scaling relations between velocity and \emph{luminous} mass are not expected for dark matter based dynamics where the masses and spatial distributions of the luminous and dark components are, a priori, independent.

\section{Observational Evidence \label{sect_obs}}

\subsection{Search for Dark Matter Particles \label{ssect_dmsearch}}

By definition, dark matter particles are supposed to interact with electromagnetic radiation either extremely weakly or not at all. Obviously, this makes a direct detection very difficult and requires the use of non-electromagnetic signatures as tracers of WIMPs \cite{gaitskell2004}. The last two decades have seen several large-scale experiments searching for characteristic recoil signals arising from the scattering of WIMPs at atomic nuclei. Those experiments are based in underground laboratories in order to minimize the non-WIMP background and use either crystals cooled to cryogenic temperatures or liquid noble gases, especially xenon, as detector materials (e.g., \cite{angloher2012,armengaud2012,aprile2012}). If an ensemble of WIMPs were to orbit the Galactic center along the solar orbit, the terrestrial motion around the sun should change the relative velocity of Earth and the WIMP ensemble during the year, leading to a characteristic annual modulation of the signal rate. In addition to direct detection methods, signatures of the decay or annihilation of WIMP particles have been searched for in cosmic $\gamma$ \cite{ackermann2012}, neutrino \cite{tanaka2011}, and antiproton \cite{cirelli2013} radiation. To date, none of the multiple experiments and observations has returned a detection of dark matter particle candidates (see \cite{cho2013b,blum2013} for very recent null results).

\subsection{Mass--velocity scaling relations \label{ssect_tully}}

\begin{figure}[t!]
\centering
\includegraphics[height=75mm,angle=-90]{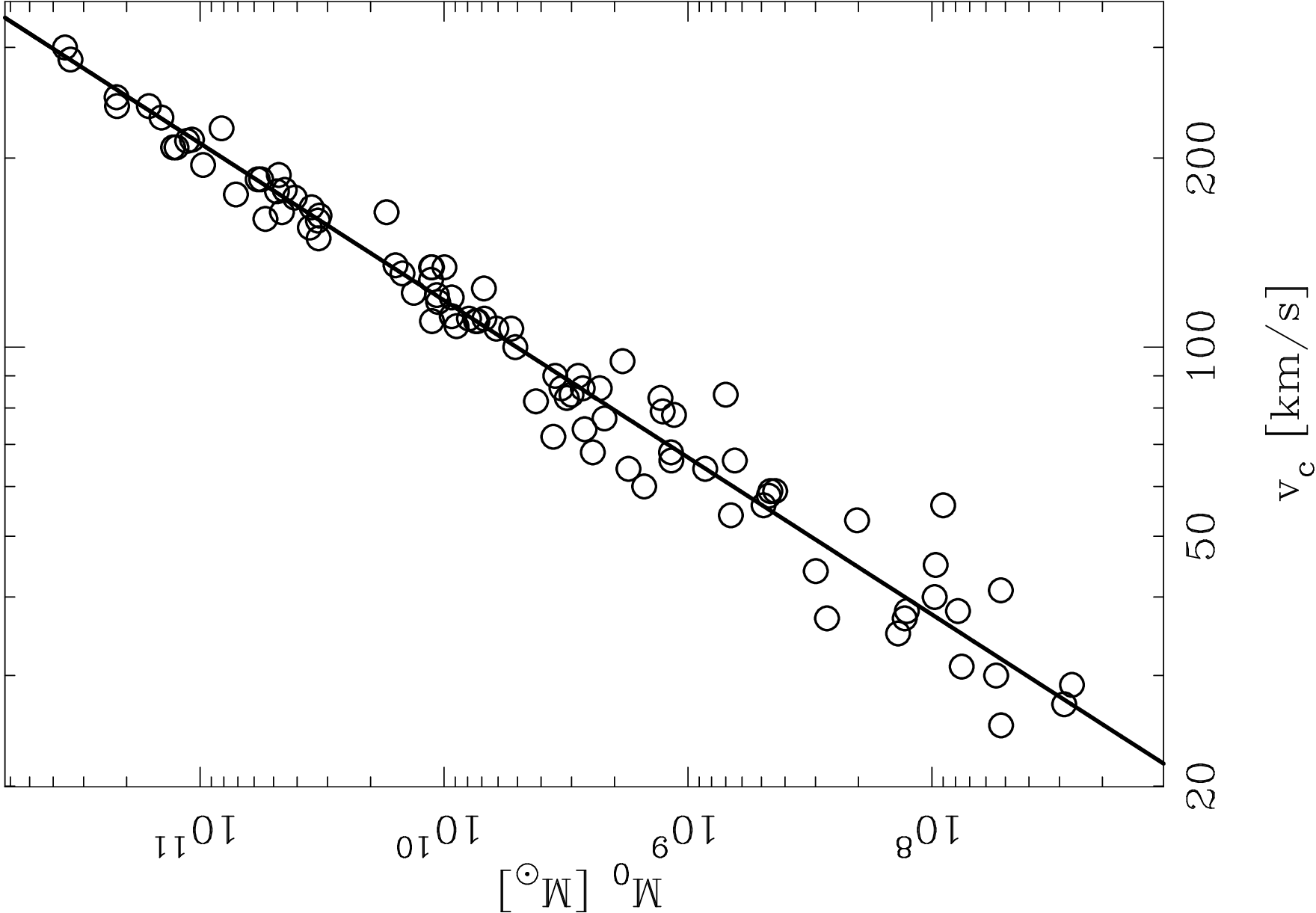}
\caption{The baryonic Tully--Fisher relation (in logarithmic representation): luminous galaxy mass $\Mo$ (in units of solar mass $\Msol$) as function of asymptotic rotation speed $\vc$. Circles denote observational values for 92 galaxies \cite{mcgaugh2005,mcgaugh2011}; the data span four orders of magnitude in mass. The straight line shows the relation $\Mo=\vc^4/(1.3\,G\,\am)$ for $\am=1.15\times10^{-10}$\mss; the factor 1.3 arises from geometry. Error bars have been omitted for clarity; the intrinsic scatter of the relation is consistent with zero. \label{fig_tully}}
\end{figure}

In 1977, \cite{tully1977} reported the discovery of an empirical correlation -- the \emph{Tully--Fisher relation} -- between the luminosity $L$ of disk galaxies and their (asymptotic) circular speed $\vc$ such that $L\propto\vc^4$. The luminosity of a galaxy is a proxy for the combined mass of its stars, $\Mstar$; deriving this mass requires knowledge of the galaxy's mass-to-light ratio $\Upsilon$. As it turns out \cite{rhee2004a,rhee2004b,mcgaugh2005}, the actual, underlying fundamental relation is found by taking into account the \emph{total} luminous (baryonic) mass of galaxies, including stars and the diffuse interstellar matter. This results in a \emph{baryonic} Tully--Fisher relation $\Mo\propto\vc^4$ holding over at least four orders of magnitude in galactic mass -- see Fig.~\ref{fig_tully} for an illustration. Comparison of the data with the scaling law (\ref{eq_vc}) leads to the relation

\begin{equation}
\label{eq_tully}
\Mo = \frac{\vc^4}{1.3\,G\,\am}
\end{equation}

\noindent
where the (approximate) geometry factor 1.3 arises from the difference between the distributions of mass in idealized homogeneous spheres and realistic disk galaxies (cf. (\ref{eq_galrot}) and \S\,2.6.1(b) of \cite{binney2008}). As illustrated in Fig.~\ref{fig_tully}, data and model are in excellent agreement for $\am\approx1.15\times10^{-10}$\mss. Taking into account the measurement errors, the intrinsic scatter of the data about the theoretical line is consistent with zero \cite{mcgaugh2005,mcgaugh2011} -- in other words, (\ref{eq_tully}) provides a \emph{complete} description of the large-scale kinematics of disk galaxies.

Already one year before the discovery of the Tully--Fisher relation, in 1976, \cite{faber1976} reported observational evidence for a strikingly similar scaling law in \emph{elliptical} galaxies. Elliptical galaxies are dominated by random stellar motions with velocity dispersion $\sigma$; \cite{faber1976} found a relation between the velocity dispersions and the luminosities of galaxies like $L\propto\sigma^4$ -- a relation known today as the \emph{Faber--Jackson relation}. In analogy to the procedure for the baryonic Tully--Fisher relation, one can assume an underlying relation $\Mo\propto\sigma^4$ and attempt a comparison to (\ref{eq_sigma}). Taking into account that $\sigma$ is actually the velocity dispersion integrated along the line of sight through an elliptical galaxy, it turns out that (\ref{eq_sigma}) indeed provides a good description of the kinematics of elliptical galaxies for $g\ll\am$ \cite{sanders2010}. 

In addition to the kinematics of galaxies, scaling laws $\Mo\propto\vc^4$ and/or $\Mo\propto\sigma^4$ have been found for

\begin{itemize}

\item[(i)] the motion of galaxies in galaxy clusters \cite{sanders2010};

\item[(ii)] the temperature $T\propto\sigma^2$ of the diffuse intra-cluster medium in galaxy clusters, i.e., $\Mo\propto T^2\propto\sigma^4$ \cite{sanders1994};

\item[(iii)] dwarf galaxies orbiting the Milky Way \cite{serra2012};

\item[(iv)] stellar motions in the outer regions (where $g<\am$) of globular star clusters \cite{hernandez2013}; and

\item[(v)] wide binary stars (where $g\lesssim\am$ for $\Mstar\approx1\Msol$ and $r\gtrsim7\,000$ astronomical units) \cite{hernandez2012}.

\end{itemize}

In total, the scaling laws (\ref{eq_vc}) and (\ref{eq_sigma}) are consistent with the kinematics of stellar systems spanning about eight orders of magnitude in size and 14 orders of magnitude in mass (see also \cite{famaey2012} for an exhaustive review); this comprises \emph{all} gravitationally bound stellar systems beyond the scale of planetary systems.

However, at least in case of disk galaxies a $\Mo\propto\vc^4$ scaling law can be found not only from (\ref{eq_vc}) but -- a priori -- also from standard Newtonian dynamics. Starting from the usual Newtonian relation $\vc^4=(G\Mo/r)^2$, we can introduce a (baryonic) \emph{surface mass density} $\Sigma_0=\Mo/(\pi r^2)$ which averages over all matter at galactocentric radii from 0 to $r$. Combining the two expressions leads to $\vc^4\propto\Mo\,\Sigma_0$; this implies a Tully--Fisher-like relation \emph{if} all disk galaxies have approximately the same surface mass density. Accordingly, we need to take a close look at the surface density of disk galaxies.

\subsection{Surface Density--Acceleration Relation \label{ssect_sda}}

\begin{figure}[t!]
\centering
\includegraphics[height=75mm,angle=-90]{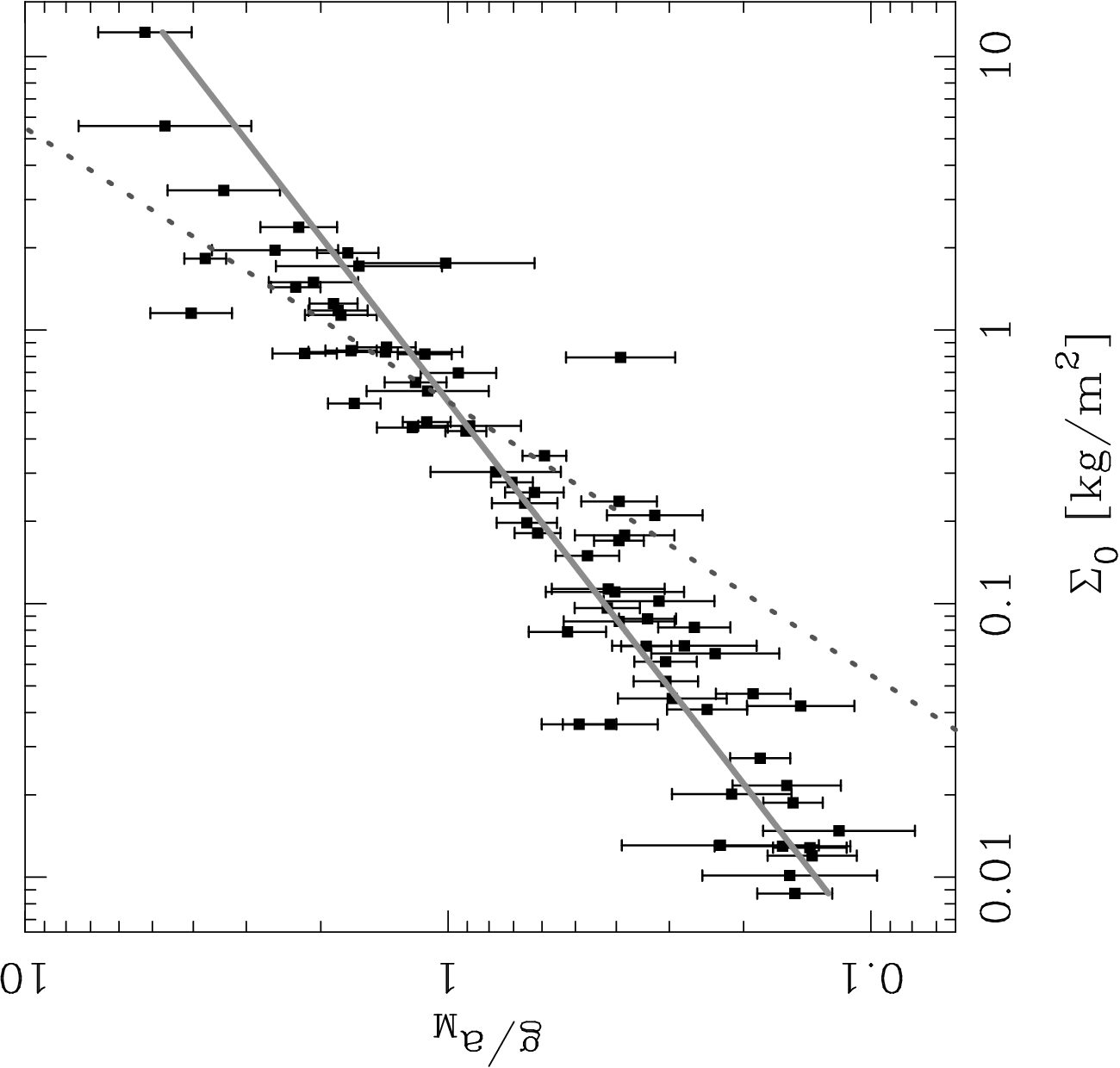}
\caption{The empirical relation between baryonic surface mass density $\Sigma_0$ and scaled gravitational acceleration $g/\am$. Data points with error bars indicate measurements from a sample of 71 disk galaxies \cite{famaey2012}. The continuous grey line corresponds to (\ref{eq_sda-mond}) with $\am=1.15\times10^{-10}$\mss; this line is purely theoretical and not a fit to the data. The dotted line denotes the relation (\ref{eq_sda-newton}) expected from Newtonian dynamics (assuming again $\am=1.15\times10^{-10}$\mss). \label{fig_sda}}
\end{figure}

Assuming the universal validity of either the Newtonian scaling law (\ref{eq_galrot}) or the modified law (\ref{eq_vc}) leads to important corollaries regarding the relation between gravitational acceleration $g$ and baryonic surface mass density $\Sigma_0=\Mo/(\pi r^2)$ of disk galaxies. In case of Newtonian dynamics, $\vc^2=G\Mo/r$. Taking this relation and (i) dividing both sides by $r$, (ii) using $g=\ac=\vc^2/r$, (iii) using the definition of $\Sigma_0$, and (iv) expressing $g$ in units of $\am$ results in

\begin{equation}
\label{eq_sda-newton}
\frac{g}{\am} = \frac{G\,\pi\,\Sigma_0}{\am} ~~ .
\end{equation}

\noindent
Ignoring the constants, this relation predicts $g \propto \Sigma_0$.

In case of modified dynamics, we start off from the relation $\vc^4=G\Mo\am$. In analogy to the procedure applied to the Newtonian case we can (i) divide both sides by $r^2$, (ii) use the definition of $\Sigma_0$, (iii) use $g=\ac=\vc^2/r$, and (iv) divide both sides by $\am^2$. Eventually, we find

\begin{equation}
\label{eq_sda-mond}
\frac{g}{\am} = \left(\frac{G\,\pi\,\Sigma_0}{\am}\right)^{1/2} ~ .
\end{equation}

\noindent
Evidently, modified dynamics predicts $g \propto \sqrt{\Sigma_0}$\,.

In Fig.~\ref{fig_sda}, I compare the predicted surface density--acceleration (SDA) relations (\ref{eq_sda-newton}, \ref{eq_sda-mond}) with observational values obtained by \cite{famaey2012} for 71 disk galaxies. For each galaxy, the radial surface density profile was (as usual in galactic astronomy) approximated as an exponential disk, i.e., $\Sigma_0(r)=\Sigma_0(0)\exp(-r/r_d)$, with a ``disk scale length'' $r_d$ as free parameter. Rotation velocities were consistently measured at $r\approx2.2r_d$, surface densities were derived from the luminous galactic mass enclosed within $r\lesssim2.2r_d$ \cite{mcgaugh2005b}; this procedure ensures that data from different galaxies can be compared in a straightforward manner. I assume here $\am=1.15\times10^{-10}$\mss\ as suggested by the baryonic Tully--Fisher relation (Fig.~\ref{fig_tully}).

The SDA observations provide three key results. Firstly, the baryonic surface mass densities of disk galaxies span over three orders of magnitude. This immediately falsifies the hypothesis that the Tully--Fisher relation arises from Newtonian dynamics combined with (approximately) constant surface densities. Secondly, the very existence of an empirical SDA relation rules out the possibility that disk galaxy dynamics is dominated by dark matter; if this were the case, we would not expect a correlation between \emph{baryonic} surface density and acceleration at all. Thirdly, the SDA data follow the theoretical line provided by MOND\footnote{The agreement with the MOND line can only be approximate because the condition $g\ll\am$ is not fulfilled for the majority of the data.} and \emph{not} the line expected from Newtonian dynamics.

\subsection{Mass Discrepancy--Acceleration Relation \label{ssect_mda}}

\begin{figure}[t!]
\centering
\includegraphics[height=75mm,angle=-90]{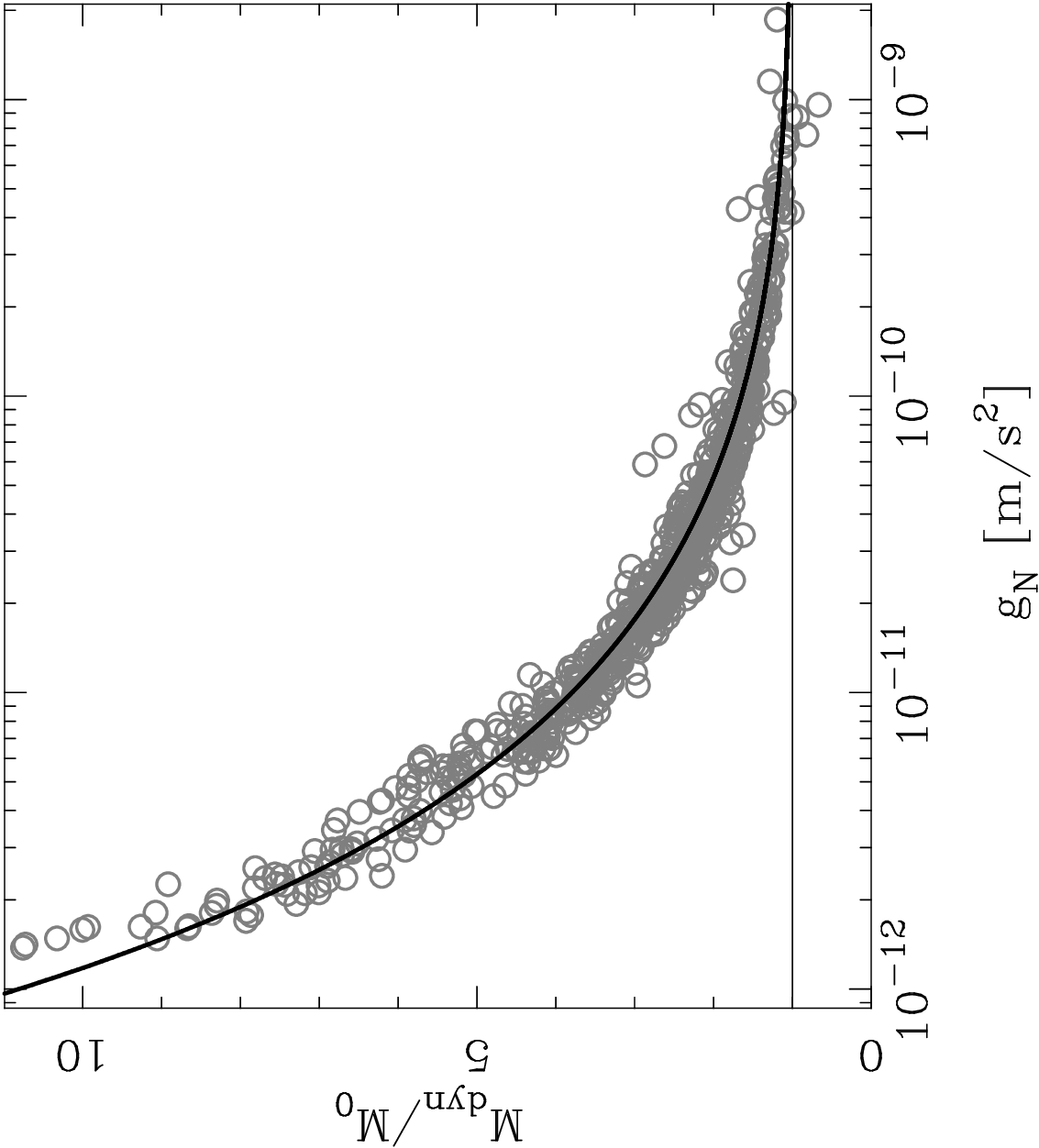}
\caption{The mass discrepancy--acceleration relation: mass discrepancy $\Md/\Mo$ as function of Newtonian gravitational acceleration $\gn=G\,\Mo/r^2$. Please note the logarithmic--linear axis scales. Grey circles denote observational values, in total 735 measurements from 60 galaxies \cite{mcgaugh2004,famaey2012,sanders2002}. The black continuous curve denotes the function (\ref{eq_mu1}) for $\am=1.06\times10^{-10}$\mss. \label{fig_mda}}
\end{figure}

Scaling relations between the luminous mass of a stellar system and the velocities of and accelerations experienced by its constituents, as discussed in \S\S\,\ref{ssect_tully} and \ref{ssect_sda}, probe the \emph{asymptotic} dynamics in the limit of weak gravitational fields. Deeper insights require an analysis of the \emph{transitional regime} from Newtonian to modified dynamics. As pointed out first by \cite{mcgaugh2004} about a decade ago, this regime can be tested by observations of stellar velocities in disk galaxies. A luminous mass $\Mo$ is derived from imaging a galaxy and summing up the masses of stars and interstellar matter; a dynamical mass $\Md$ is calculated from stellar velocities (derived from spectroscopic measurements) via (\ref{eq_galrot}). The empirical mass discrepancy (MD) $\Md/\Mo$ can be analyzed as function of galactocentric radius $r$, angular frequency, and Newtonian acceleration $\gn=G\Mo/r^2$ of stars orbiting the galactic center. Empirically, the MD is \emph{uncorrelated} with radius or angular frequency and strongly \emph{anti-correlated} with acceleration; this behavior is illustrated in Fig.~\ref{fig_mda}. Evidently, the mass discrepancy is a function of gravitational field strength $g=\vc^2/r=G\Md/r^2=\gn(\Md/\Mo)$. Notably, the mass discrepancy--acceleration (MDA) relation is \emph{universal} because all disk galaxies studied -- 60 galaxies spanning about two orders of magnitude in size in case of the data set shown in Fig.~\ref{fig_mda} -- follow the \emph{same} empirical curve \cite{mcgaugh2004,famaey2012,sanders2002}.

The observed MDA relation naturally suggests a description using the model (\ref{eq_mu}); by construction, $\mu(x)=\gn/g=\Mo/\Md$ (i.e., the inverse of the mass discrepancy). The transition function $\mu(x)$ does not follow from theory; a priori, various -- empirically motivated -- choices for the functional form of $\mu(x)$ are available \cite{milgrom1983b,famaey2005}. A good description of the MDA data is provided \cite{trippe2013c} by the choice

\begin{equation}
\label{eq_mu1}
\mu(x) = \frac{x}{1 + x} ~~~~~ {\rm with} ~~~~~ x = \frac{g}{\am} ~ ;
\end{equation}

\noindent
a possible physical motivation for this form of $\mu(x)$ will be discussed briefly in \S\,\ref{sect_discuss}. As illustrated in Fig.~\ref{fig_mda}, model and data are in very good agreement for $\am=1.06\times10^{-10}$\mss; I note that here the best-fit value for $\am$ depends on the model chosen for $\mu(x)$. Within a combined, largely systematic relative uncertainty of about 10\%, the value for $\am$ found from use of (\ref{eq_mu1}) is identical to the one derived independently from the baryonic Tully--Fisher relation (cf. Fig.~\ref{fig_tully}).

\subsection{Local Coupling of Mass and Kinematics \label{ssect_renzo}}

In \S\S\,\ref{ssect_tully} and \ref{ssect_sda}, I discussed \emph{global} scaling relations between the luminous mass and the rotation speed of disk galaxies -- ``global'' in the sense that the scaling relations discussed there are obtained from averaging over entire galaxies. An important amendment to this approach is provided by taking a close look at the corresponding \emph{local} scaling relations obtained from spatially resolving the luminous mass distributions and kinematics of individual galaxies. Observations of baryonic mass profiles -- luminous mass as function of galactocentric radius -- and rotation curves -- rotation speed as function of galactocentric radius -- on a galaxy-by-galaxy base unveil a relation that has since become known as \emph{Renzo's rule} \cite{sancisi2004}: ``For any feature in the luminosity profile there is a corresponding feature in the rotation curve and vice versa''. A more recent and more detailed study \cite{swaters2012} based on a sample of 43 disk galaxies finds that ``the observed rotation curves of disk galaxies [...] can be fitted remarkably well simply by scaling up the contributions of the stellar and H\,{\sc i} disks'' -- in other words, at any galactocentric radius the dynamical mass is completely determined by the enclosed luminous mass. Quantitatively, the best description of galactic rotation curves is achieved by (\ref{eq_mu1}) with $\am=(1.2\pm0.3)\times10^{-10}$\mss\ \cite{gentile2011}. Such a behavior is self-evident in the context of modified gravity models (where the luminous mass \emph{is} the dynamical mass when taking into account the modified law of gravity); it is extremely difficult to understand when starting off from the assumption of a separate dark mass distribution around a galaxy.

\subsection{Dwarf Galaxies \label{ssect_dwarfs}}

The modern textbook picture of galaxies \cite{binney1998,sparke2007} usually assumes that most (crudely 50--90\%) of the total mass of any galaxy (regardless of the type of galaxy) is stored in a spherical dark matter halo that extends well beyond the luminous components of the galaxy. A key prediction of this picture is the formation of a large number of small ``dwarf'' satellite galaxies -- due to ``condensation'' of luminous matter in local dark matter concentrations within the dark halo -- distributed isotropically around the main galaxy. This is actually not observed: the number of satellite galaxies observed around galaxies is substantially -- by more than one order of magnitude -- smaller than expected (a result known as the \emph{missing satellite problem}). Furthermore, the distribution of these satellites is not isotropic; the satellites of the Milky Way and a few other galaxies are located in planes that coincide with the planes of past galaxy--galaxy collisions -- which indicates that all these satellites are ``tidal dwarf galaxies'' formed via tidal interactions between galaxies. For a recent exhaustive review, see \cite{kroupa2012} and references therein.

\subsection{Colliding Galaxy Clusters \label{ssect_galcoll}}

On the one hand, modified laws of gravity demand that the luminous matter of any stellar system traces the total dynamical mass. On the other hand, dynamical models based on dark matter do not comprise this restriction; the spatial distribution of the total dynamical mass does not necessarily follow the luminous matter. Accordingly, a decision in favor of dark matter based models could be enforced by observations of astronomical objects where dynamical mass and luminous mass are spatially separate. A key test is provided by clusters of galaxies that are colliding or, more appropriately speaking, passing through each other. Due to their very small effective cross section, the individual galaxies are de facto collisionless and just pass each other without interaction. The same can be assumed for the (inter)galactic dark matter halos -- if those actually exist. The intergalactic cluster gas -- that comprises the overwhelming fraction of the total luminous mass of a galaxy cluster --  however experiences collisional ram pressure and lags behind the galaxies and the dark matter distributions. Observationally, the distribution of the luminous mass is -- essentially -- traced by the X-ray emission from the intracluster gas; the distribution of the gravitating dynamical mass is derived from the gravitational lensing of background sources.

The first system of colliding clusters that has been studied in detail is the ``Bullet Cluster'' 1E0657-56 \cite{barrena2002}. Comparison of the distributions of intracluster gas and dynamical mass shows a clear spatial separation -- most of the dynamical mass indeed appears to be stored in a non-luminous mass component, and accordingly, this result was interpreted as a triumph of the dark matter paradigm over its competitors. Ironically, more recent studies of the kinematics of the Bullet Cluster \cite{lee2010} have turned around this interpretation completely: the relative velocities of the colliding galaxy clusters are far too high ($\approx$3000\,km/s compared to $\lesssim$1800\,km/s expected theoretically) to be understood by dark matter driven dynamics. Using results from cosmological numerical simulations for reference, \cite{lee2010} conclude that the probability for the standard dark matter paradigm being correct is less than $10^{-9}$. At this point it is important to note that the initial interpretation of the mass distributions within the Bullet Cluster is based on the presumption that (essentially) the entire baryonic intracluster medium in galaxy clusters is observable as X-ray luminous gas. This is probably not the case: analyses of the baryonic mass content of galaxy clusters conclude that crudely half of the total baryonic matter seems to have been missed by observations as yet \cite{shull2012,fukugita2004}. Furthermore, a second system of colliding galaxy clusters has become available for analysis: the ``Train Wreck Cluster'' A520 \cite{markevitch2005}. In A520, observations find that the spatial distributions of dynamical and luminous mass coincide \cite{jee2012}.

\section{Discussion \label{sect_discuss}}

Ever since its discovery, the missing mass problem has been a challenge to our understanding of the physics of the cosmos. Historically, the standard approach has been the postulate of cold dark matter (\S\,\ref{sect_dm}; cf. also \cite{ostriker1973,einasto1974}), and, initially, for a very good reason: else than a modified law of gravity, the assumption of dark matter does not per se require ``new physics'' -- as long as it can be reconciled with the standard model of particle physics. As I outlined in \S\,\ref{ssect_wimp}, Occam's razor is no longer in favor of dark matter: at least since the failed searches for MACHOs in the 2000s, models based on dark matter inevitably require physics beyond the standard model of particle physics -- \emph{both} dark matter and modified gravity now require new physics. Else than usually, and implicitly, assumed (or hoped for), the two approaches are not mutually exclusive; this is important for the discussion of structure on cosmological scales.

A review of the observational evidence provided in \S\,\ref{sect_obs} leads to conclusions that fall into either of two categories:

\para{I. Failures of the dark matter paradigm.}  The most evident problem of the dark matter paradigm is the fact that a particle with the required physical properties is not known; to date, none of the multiple direct and indirect searches for dark matter particles has returned any viable candidate (\S\,\ref{ssect_dmsearch}). In order to understand the kinematics of galaxies it is necessary to assume that galaxies are embedded in extended, more or less spherical, dark matter halos with mass density profiles $\rho\propto r^{-2}$ over a sufficient range of galactocentric radii $r$. Those models imply the presence of large numbers of isotropically distributed dwarf satellite galaxies -- which are not observed (\S\,\ref{ssect_dwarfs}) -- as well as certain limits on the collision velocities of galaxy clusters -- which are in disagreement with observations (\S\,\ref{ssect_galcoll}).

Taking a more careful look at galactic kinematics, it turns out that the presence of dark matter halos with $\rho\propto r^{-2}$ can explain the asymptotic flattening of galactic rotation curves ($\vc(r)\approx const$ at large $r$) -- but nothing else. Indeed, \emph{none} of the relevant empirical scaling relations of galactic kinematics -- notably the Tully--Fisher and Faber--Jackson relations (\S\,\ref{ssect_tully}), the surface density--acceleration (\S\,\ref{ssect_sda}) and mass discrepancy--acceleration (\S\,\ref{ssect_mda}) relations, and Renzo's rule (\S\,\ref{ssect_renzo}) -- follow from dark matter models a priori. Explaining the baryonic Tully--Fisher relation requires a ``fine tuning'' of dark and luminous matter such that for all galaxies the \emph{total} surface mass density $\Sigma=\Sigma_0+\Sigma_{\rm DM}$ (with $\Sigma_{\rm DM}$ being the dark matter surface density) is approximately the same despite the fact that $\Sigma_0$ varies over three orders of magnitude. Arguably the only realistic way to achieve $\Sigma\approx const$ is the assumption that $\Sigma_{\rm DM}\gg\Sigma_0$ for all galaxies plus $\Sigma_{\rm DM}\approx const$. However, this ansatz is contradicted by the empirical SDA relation which demonstrates that galactic dynamics is controlled by the \emph{luminous} matter, implying $\Sigma_{\rm DM}\ll\Sigma_0$. To date, the necessary fine tunings between (i) dark and luminous surface densities, as well as (ii) between dark and luminous mass profiles within galaxies (Renzo's rule) -- also referred to as a dual ``conspiracy between luminous and dark matter'' \cite{rhee2004a} -- are unexplained.

\para{II. Successes of modified dynamics.}  The three predictions P1--3 provided by Modified Newtonian Dynamics (\S\,\ref{sect_mog}) can be compared to observations in a straightforward manner. Prediction P1 states that luminous and dynamical mass are identical when applying an appropriate scaling. This is observed in individual galaxies as Renzo's rule (\S\,\ref{ssect_renzo}) as well as in at least one galaxy cluster, A520 (\S\,\ref{ssect_galcoll}). Prediction P2 demands a universal scaling of mass discrepancies according to (\ref{eq_mu}) and, indeed, observations find a universal mass discrepancy--acceleration relation of disk galaxies (\S\,\ref{ssect_mda}). Prediction P3 states that, for $g\ll\am$, stellar velocities and luminous masses are related according to (\ref{eq_vc}, \ref{eq_sigma}). Comparisons to observations show that P3 naturally provides for the baryonic Tully--Fisher and Faber--Jackson relations on all scales (\S\,\ref{ssect_tully}) as well as for the surface density--acceleration relation of disk galaxies (\S\,\ref{ssect_sda}). Remarkably, the scaling laws (\ref{eq_vc}, \ref{eq_sigma}) are consistent with the kinematics of stellar systems spanning eight orders of magnitude in size and 14 orders of magnitude in mass, thus covering \emph{all} gravitationally bound stellar systems beyond the scale of planetary systems. I emphasize that the statements P1--3 are indeed predictions: even though Milgrom constructed (\ref{eq_mu}) based on his knowledge of flat rotation curves and the Tully--Fisher relation, most of the relevant observations, notably the \emph{baryonic} Tully--Fisher and Faber--Jackson relations, Renzo's rule, and the SDA and MDA relations, have been achieved only as late as two decades after Milgrom's proposal. Last but not least, MOND is not only physically more successful than dark-matter driven dynamics but also technically simpler: modeling the dynamics of any given galaxy with MOND requires one free parameter, the mass-to-light ratio $\Upsilon$ (cf. especially \S\S\,\ref{ssect_tully}, \ref{ssect_mda}). When fitting a dark matter halo to a given galaxy, one needs at least three parameters: $\Upsilon$ and (at least) the scaling parameters $\rho_0$ and $r_0$ (cf. \S\,\ref{sect_dm}). 

\para{\indent}Regarding the combined evidence, it is rather obvious that the missing mass problem is treated much better by a modified law of gravity than by postulating dark matter. This being said, it is also clear that MOND is incomplete: it is a \emph{law} of gravity but not yet a \emph{theory} of gravity -- meaning it does not provide the underlying physical mechanism a priori. A key question to be addressed is the physical role of Milgrom's constant; empirically (\S\,\ref{sect_obs}), $\am=(1.1\pm0.1)\times10^{-10}$\mss, with the error being largely systematic. Coincidentally, $\am=c\,H_0/2\pi$ within errors, with $c$ being the speed of light and $H_0\approx70$\,km\,s$^{-1}$\,Mpc$^{-1}$ being Hubble's constant \cite{lee2012}; this might indicate an as yet uncovered connection between galactic dynamics and cosmology. Over the last decades, there have been multiple approaches toward theories of gravity beyond Einstein's GR, and the following is a brief, incomplete overview over six important examples:

\epara{Tensor--vector--scalar theory (TeVeS)} was derived by Jacob Bekenstein as a relativistic gravitation theory of MOND \cite{bekenstein2004,bekenstein2006}. In TeVeS, gravitation is mediated by three dynamical gravitational fields: the Einstein metric tensor $g_{\mu\nu}$, a timelike 4-vector field $U_{\alpha}$ obeying $g^{\mu\nu}U_{\mu}U_{\nu}=-1$, and a scalar field $\phi$ (using here the usual Einsteinian index conventions). By construction, TeVeS is consistent with general relativity and the relevant limits, Newtonian dynamics (for $g\gg\am$) and MOND (for $g\ll\am$).

\epara{$f(\R)$ gravity} results from theories aimed at generalizations of the Lagrangian $\L$ in the Einstein--Hilbert action of general relativity (see \cite{sotiriou2010,defelice2010,nojiri2011} and references therein for reviews). In standard GR, $\L\propto\R$ with $\R$ being the Ricci scalar. In a generalized formulation, this expression becomes $\L\propto f(\R)$ with $f(\R)$ being a general function of $\R$; a simple example is $f(\R)=\R^n$ with $n$ being a real number. $f(\R)$ theories have been motivated mostly by cosmology, especially the \emph{cosmic inflation} and \emph{dark energy} problems, but have also found application to the missing mass problem (e.g., \cite{capozziello2004}).\footnote{Technically, already the introduction of a cosmological constant $\Lambda$ into Einstein's field equations leads to an $f(\R)$ gravity with $f(\R)=\R-2\Lambda$ \cite{defelice2010}.}

\epara{Massive gravity} follows from the assumption (motivated by quantum field theories) that gravitation is mediated by virtual particles, \emph{gravitons}, with non-zero mass. This possibility was first pointed out by Fierz and Pauli in 1939 \cite{fierz1939} and has been studied extensively especially in view of potential applications to cosmology (see \cite{goldhaber2010,hinterbichler2012} for reviews). Massive gravity provides an intuitive approach to the missing mass problem: assuming a luminous mass $\Mo$ and no graviton--graviton interactions, the emitted gravitons form a halo with mass density profile $\rho(r)=\beta\,\Mo\,r^{-2}$ with $r$ being the radial distance and $\beta$ being a scaling factor. Using $\beta=\am/(4\pi\vc^2)$, with $\vc$ being the circular speed of a test mass orbiting $\Mo$, and integrating over $r$ leads to (\ref{eq_mu1}) \cite{trippe2013c,trippe2013a,trippe2013b}.

\epara{Non-local gravity} is based on the perception that the postulate of locality in the special theory of relativity is only approximately correct for realistic accelerated observers. This suggests the construction of a non-local general theory of relativity wherein the gravitational field is local but satisfies non-local integro-differential equations. In the Newtonian limit one finds a modified Poisson equation and from this a gravitational point mass potential which comprises an extra term that mimics an additional ``dark'' mass component \cite{mashhoon2008,hehl2009,blome2010}.

\epara{Scalar--tensor--vector gravity}, also referred to as Modified Gravity (MOG), modifies Einsteinian (tensor) GR by adding one massive vector field and two scalar fields as mediators of gravitation. The theory contains two free parameters that can be constrained empirically by comparison to galaxy rotation curves. Once this is achieved, the only free parameter remaining for the description of any given galaxy is the galactic mass-to-light ratio $\Upsilon$ \cite{moffat2006,moffat2013}.

\epara{Conformal gravity} follows from the insight that (i) the Einstein equations are not the only possible field equations of GR, and (ii) the Einstein--Hilbert action is not the only possible action of GR (cf. also the case of $f(\R)$ gravity). Accordingly, it is possible to derive a theory of gravitation that keeps the coordinate invariance and equivalence principle structure of general relativity but adds a local conformal invariance in which the action is invariant under a specific local transformation of the metric. This results in additional \emph{universal} gravitational potential terms of linear and higher orders. The modified gravitational potential is supposed to mimic the effect of dark matter in galaxies \cite{mannheim2012a,mannheim2012b,yoon2013}.

\epara{\indent}Any theory of modified gravity necessarily needs to comprise the -- empirically well established -- MOND laws as well as non-dynamical signatures of gravity, notably (i) gravitational lensing and (ii) gravitational waves. Studies assuming TeVeS as the underlying theory of gravity find good agreement with observations of gravitational lensing \cite{shan2008,chiu2011,tian2013}. An example for a theory naturally including gravitational waves is provided by massive gravity; as already pointed out by \cite{fierz1939}, in massive gravity the GR equations of gravitational waves are recovered in the limiting case of vanishing graviton mass.

The discussion of the missing mass problem and modified dynamics has implications for, and has to be consistent with, cosmology. Modern cosmological models are constrained best by observations of the angular power spectrum of the cosmic microwave background (CMB) radiation \cite{ade2013,penzias1965,larson2011}; the current concordance cosmology is the $\Lambda$ Cold Dark Matter ($\Lambda$CDM) model, with $\Lambda$ denoting the cosmological constant \cite{bahcall1999}. Using a total of six fit parameters, the $\Lambda$CDM model indeed provides a very good description of the CMB power spectrum \cite{ade2013}. This success comes at a high prize however: it requires the assumption that only 5\% of the mass/energy content of the universe are provided by ``ordinary'' luminous matter; 27\% are provided by non-baryonic cold dark matter, and the remaining 68\% by \emph{dark energy}, a dark fluid generating negative gravitational pressure and thus causing an accelerated expansion of the universe. Whereas the $\Lambda$CDM model is \emph{technically} successful, the fact that it requires 95\% of the mass/energy content to be provided by exotic dark components makes it \emph{physically} dubious.\footnote{Quoting the more drastic wording by \cite{cho2013a}: ``According to {\sl Planck} [the CMB space observatory], the universe consists of 4.9\% ordinary matter, 26.8\% mysterious dark matter whose gravity holds the galaxies together, and 68.3\% weird, space-stretching dark energy.'' In essence: when assuming a $\Lambda$CDM cosmology, 95\% of the mass/energy content of the universe are ``mysterious'' and/or ``weird''.} More importantly, the $\Lambda$CDM model is not the only possible description of the CMB power spectrum: the data can be explained equally well by assuming Modified Newtonian Dynamics plus \emph{leptonic, hot} dark matter -- distributed over the spatial scales of galaxy clusters and larger -- composed of neutrinos with masses on the order of few electron volts \cite{angus2011}. Furthermore, and very recently, new tools have been developed for testing the impact of modified laws of gravity on the large-scale structure of cosmic matter via numerical simulations. First results indicate that the observational signatures of a combination of $f(\R)$ gravity and massive neutrinos are indistinguishable from those of $\Lambda$CDM models \cite{puchwein2013,baldi2013}. I note however that these results do not yet imply the discovery of a cosmology based on modified gravity; such a cosmology needs to be derived from first principles out of a modified theory of gravity -- which has not been achieved yet.

A complication arises for the dynamics of X-ray bright groups and clusters of galaxies: even when assuming MOND, the amount of luminous mass observed is too small -- by factors on the order of two -- to explain the dynamics \cite{angus2008,famaey2012}. This discrepancy is especially pronounced in the central regions of those systems. A consistent dynamical description requires (i) a modified theory of gravity that leads to scaling laws more complex than (\ref{eq_mu}) and (\ref{eq_sigma}) on scales of groups of galaxies (i.e., hundreds of kiloparsecs); (ii) cold \emph{baryonic} dark matter (CBDM) presumably in the form of cold (few Kelvin) compact gas clouds \cite{milgrom2008}; or (iii) the presence of massive (preferred masses being about 11\,eV) \emph{sterile} neutrinos \cite{angus2009}.

Summarizing over the multiple lines of evidence -- from stellar dynamics to cosmology -- it is difficult to resist the impression that a solution of the missing mass problem requires a modified law of gravity; the standard dark matter postulate seems to be more and more disfavored by observations. The latter statement is specifically aimed at ``naive'' dynamical models based solely on non-baryonic cold dark matter; certain combinations of modified gravity and standard-model dark matter -- especially massive neutrinos and/or CBDM -- are in good agreement with cosmological observations and might actually be required for understanding the dynamics of galaxy clusters. Such a paradigm change also has implications for particle physics where the presumed need for exotic dark matter particles in astronomy has become a textbook justification for a supersymmetric extension of the standard model of particle physics (e.g., \cite{martin2011,griffith2008}). Evidently, our current understanding of the framework required for a modified theory of gravity is sketchy at best and needs substantial additional work -- a lot has been done already, but even more is left to do.

\section{Summary and Conclusions \label{sect_conclude}}

Since the 1930s, astronomical observations have consistently found evidence for a systematic missing mass problem: assuming the validity of Newtonian dynamics on astronomical scales, the dynamical, gravitating masses of galaxies required to explain their kinematics exceed their luminous masses by up to one order of magnitude. Traditionally, this effect has been explained by assuming galaxies (and clusters of galaxies) to be embedded within halos of non-baryonic, cold, and electromagnetically dark matter. Within the last decade, astronomical observations of multiple star systems, star clusters, galaxies, and galaxy clusters have found various kinematic scaling relations that shed new light on the missing mass problem: the baryonic Tully--Fisher and Faber--Jackson relations, ``Renzo's rule'', and the surface density--acceleration and mass discrepancy--acceleration relations. 

On the one hand, the traditional dark matter picture has serious difficulties to explain the observations: \emph{none} of the relevant kinematic scaling relations follows from the assumption of dark matter halos; assuming stellar dynamics to be controlled by dark matter requires an extremely implausible ``fine tuning'' between the spatial distributions of dark and luminous matter in galaxies. To date, dark matter particles have not been detected despite massive experimental efforts. Furthermore, the presence of dark matter halos is incompatible with distributions of dwarf galaxies around ``mother'' galaxies as well as the kinematics of colliding galaxy clusters.

On the other hand, \emph{all} observations are explained naturally in the frame of Modified Newtonian Dynamics (MOND) which assumes a characteristic re-scaling of Newtonian gravitational acceleration as function of acceleration (field strength). Most of the aforementioned kinematic scaling relations were predicted successfully by MOND as early as about two decades before they were actually observed. MOND is also found to be consistent with cosmological observations, most notably the angular power spectrum of the cosmic microwave background radiation.

Regarding the combined evidence, it becomes more and more obvious that the solution for the missing mass problem is to be found in a modified theory of gravity that comprises the MOND laws. Despite multiple attempts on theories of gravity beyond Einsteinian general relativity, a consistent picture has not yet emerged and is left as work still to be done.

\paragraph{Acknowledgments:} This work made use of the galactic dynamics data base provided by \name{Stacy S. McGaugh} at Case Western Reserve University, Cleveland (Ohio),\footnote{\url{http://astroweb.case.edu/ssm/data/}} and of the software package \name{DPUSER} developed and maintained by \name{Thomas Ott} at MPE Garching.\footnote{\url{http://www.mpe.mpg.de/~ott/dpuser/dpuser.html}} I acknowledge financial support from the Korean National Research Foundation (NRF) via Basic Research Grant 2012R1A1A2041387. Last but not least, I am grateful to four anonymous referees for valuable suggestions and comments.

\end{document}